%% file: Orbital-collapse-PRL-v0.tex
\documentclass[aps,prl,twocolumn,10pt,reprint,floatfix,superscriptaddress]{revtex4-2}
\usepackage{amsfonts,amsmath,amssymb,bm,color}
\usepackage{hyperref}
\usepackage{graphicx}
\usepackage[utf8]{inputenc}
\UseRawInputEncoding
\usepackage{orcidlink}

\usepackage{color}
\usepackage{siunitx}

\begin{document}
\newlength{\figwidth}
\setlength{\figwidth}{0.45\textwidth}
\setlength{\figwidth}{7.8cm}

\title{Orbital Collapse in Exotic Atoms and Its Effect on Dynamics}
\author{X. M. Tong \orcidlink{0000-0003-4898-3491} }\email{tong.xiaomin.ga@u.tsukuba.ac.jp}
\affiliation{
Center for Computational Sciences, University of Tsukuba, Tsukuba, Ibaraki 305-8573, Japan}

\author{K. T\H ok\'esi \orcidlink{0000-0001-8772-8472}}
\affiliation{HUN-REN Institute for Nuclear Research (ATOMKI), 4026 Debrecen, Hungary}

\author{D. Kato \orcidlink{0000-0002-5302-073X} }
\affiliation{National Institute for Fusion Science (NIFS), Toki, Gifu 509-5292, Japan}
\affiliation{Interdisciplinary Graduate School of Engineering Sciences, Kyushu University, Kasuga, Fukuoka 816-8580, Japan}

\author{T. Okumura \orcidlink{0000-0002-3037-6573}}
\affiliation{Department of Chemistry, Tokyo Metropolitan University, Hachioji, Tokyo 192-0397, Japan}

\author{S. Okada \orcidlink{0000-0002-5349-8549}}
\affiliation{National Institute for Fusion Science (NIFS), Toki, Gifu 509-5292, Japan}
\affiliation{Department of Mathematical and Physical Sciences, Chubu University, Kasugai, Aichi 487-8501, Japan}

\author{T. Azuma \orcidlink{0000-0002-6416-1212}}\email{toshiyuki-azuma@riken.jp}
\affiliation{Atomic, Molecular and Optical Physics Laboratory, RIKEN, Wako, Saitama 351-0198, Japan}

\begin{abstract}
We study the energy structures of muonic Ar atoms and find the muon orbital collapses at a critical angular momentum $l_c$ using density-functional theory (DFT).  The $l_c$ may provide an upper limit for the muon-captured states in muon-Ar  collisions. We confirm the  existence of this upper limit by calculating the state-specified capture probability using the time-dependent Schr\"odinger equation (TDSE) and a classical trajectory Monte Carlo (CTMC) methods with the single-active-particle approximation.  Modifying the mapping between the classical binding energy and the principal quantum number led to a reasonable agreement in the state-specified muon capture probabilities obtained by the TDSE and CTMC methods. We also propose a simple method to estimate $l_c$ for exotic noble atoms from atomic model potentials. The estimated values agree with those calculated by DFT.
\end{abstract}
\maketitle

Fermi predicted the atomic orbital collapse based on the Thomas-Fermi model \cite{Fermi28}. 
The phenomenon occurs when the $4f$ orbital has a higher energy than the $6s$ orbital, and it arises mainly for high-$Z(>50)$ atoms. The $f$-orbital collapse results in the $f$-block \cite{Mayer41a,Jensen82} elements in the periodic table. The $5g$ orbital collapse was predicted for $Z>121$ \cite{Griffin69,Tupitsyn24}. The $3d$ orbital also shows weak collapse  for $Z>20$ \cite{Connerade20}, since  the $3d$ orbital  is higher in energy than the $4s$ orbital \cite{Griffin69} for transition metals. We define the electron in the collapsed orbital as an active electron. 
The physical origin of orbital collapse is  the potential barrier formed by  the interactions of the active electron with other electrons (electron screening) and the nucleus (Coulomb interaction), and the centrifugal potential of the active electron. 
Two  potential wells (inner and outer) are formed by the potential barrier. The inner potential well is shallow for low-$Z$  atoms, and its depth increases with $Z$. At a certain $Z$, the well can  hold a bound state, causing  the orbital collapse. Therefore, orbital collapse is not observed in  elements lighter than Ar.

We theoretically  identify a new type of  orbital collapse:  the orbital collapse of a bound negatively charged particle in an exotic atom. Exotic atoms  are atoms in which one electron is replaced by a  negatively charged heavy particle, such as a muon, pion, kaon, or antiproton \cite{Fermi47}. 
Typically, the heavy particle is first captured into a highly excited state, which then cascades to lower excited states through Auger or radiative decay  before reaching the ground state \cite{Hufner66,Akylas78}. Due to its  mass, the heavy particle approaches the nucleus and probes the nuclear structure, acting as a bridge between atomic and nuclear physics \cite{Measday01,Gorringe15,Knecht20,Valuev22}.  Even when the particle is in a lower excited state, it  is closer to the nucleus than the inner-shell electrons. Therefore, the x rays emitted during the cascade  contain information  on the bound state quantum electrodynamics (QED) \cite{Paul21,Okumura23}. High-precision laser spectroscopy has been used to study the properties of pion \cite{Hori20}. Antihydrogen  is a good candidate to test the charge-parity-time reversal (CPT) theorem \cite{Andresen11}, which is a  fundamental symmetry of nature \cite{Kostelecky11}.  A heavy particle could be captured into a very highly excited state if it collided with an excited atom. For example, the collision of  an antiproton with an excited Xe atom could cause it to be  captured into a Rydberg state with a principal quantum number close to a thousand. Such a Rydberg state of heavy particles  is applicable to study quantum to classical transitions \cite{Yoshida00,Le05,Reinhold05a}.

The initial capture state of a heavy particle is crucial to the analysis of subsequent processes. 
However, the initial state-specified capture probability remains poorly understood (see the review in Ref.~\cite{Hartmann93}). The state-specified protonium formation, a typical Coulomb three-body problem, has been studied by solving the time-dependent Schr\"odinger equation (TDSE) \cite{Sakimoto02,Tong06c} and by a classical trajectory Monte Carlo (CTMC) method \cite{Abrines66,Olson77,Cohen99}. 
In principle, the capture processes of a many-electron atom can be studied by CTMC. However, it requires substantial  computational resources. Therefore, one has to practically study the problem using various approximations for many-electron systems.

Stimulated by recent muon experiments on Fe metal \cite{Okumura21} and Ar gas \cite{Okumura24}, we investigated the structures of muonic atoms and the muon initial-captured states. A muon is about 207 times heavier than an electron. For H atoms, a muon of mass $m_\mu$ is mainly captured into a state with a principle quantum number $n_\mu=\sqrt{m_\mu}=14$  based on a mass scaling relation \cite{Cohen83,Tong07}. 
We use atomic units ($m_e=\hbar=e=1$)  in this work unless  stated otherwise. In contrast,  our analysis of  the electronic $K$ x rays emitted from muonic Ar atoms \cite{Okumura24} shows the initial capture states to be much higher.  This finding was the main motivation for the present investigation. We consider only the $3p$ electron as an active electron, despite other orbitals (e.g.,  $3s$ and $2p$) also contributing to the capture process with lower probabilities. 

\begin{figure}[tb]
\centering{\includegraphics[width=\figwidth]{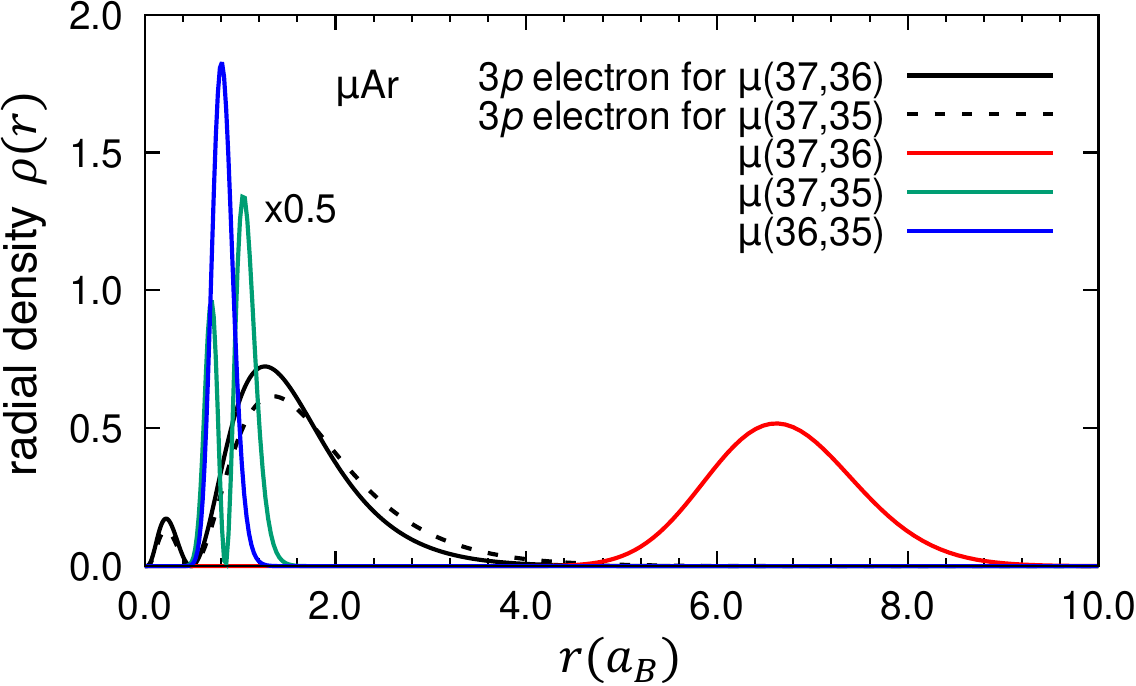} }
\caption{Radial density $\rho(r)$ of muon orbitals of $(n_\mu,l_\mu) = $ (37,36), (37,35), (36,35) and the $3p$ electron orbital (muon in (37,36) solid line,  or (37,35) dashed line) for $\mu$Ar  calculated by DFT. The radial densities of muon in (37,36) and (36,35) states are scaled down by 0.5.
$a_B$ is the Bohr radius for electron. \label{fig:1}}
\end{figure}

The capture probability is determined by the overlap of the muon orbital with the $3p$ electron orbital, the interaction time between the two particles, and the statistical weight of $(2l_\mu+1)$ with $l_\mu$ the muon angular momentum. The overlap is associated with  the mean radius of the muon-captured state and the interaction time  relates to the energy of the ejected electron and the muon binding energy. Figure~\ref{fig:1} shows the radial densities of the muon orbitals and the $3p$ electron orbital calculated by  self-interaction-free density functional theory (DFT)  \cite{Tong23}.
We found that the muon orbital $(n_\mu,l_\mu)=(37,36)$ is pushed to the outer side of  the  $3p$ electron orbital, with almost no overlap between them. For the $(37,35)$ and $(36,35)$ states, the muon orbitals shrink to the inner region and largely overlap with the $3p$ electron orbital.  This behavior is very similar to the well-known atomic orbital collapse in high-$Z$ atoms.

Orbital collapse means that the orbital wavefunction changes dramatically for a circular state $(n,l=n-1)$ if $l$ changes from $l+1$ to $l$. 
Similarly to atomic orbital collapse, the physical origin is the potential barrier formed due to the competition of the interactions among three types of charged particle: muon, electron, and nucleus. The muon-nucleus interaction and the screening of the electrons are approximated by atomic model potentials. We use two kinds of model potential to describe the active particle (muon or electron)  interacting with the nucleus and its surrounding electrons.  One, which we call here model potential 1, is a six-parameter potential in the following form \cite{Tong05c} 
\begin{eqnarray}\label{eq:mp1}
V(r) = -\frac{ 1 + a_1 e^{-a_2 r}+a_3 r e^{-a_4 r}+a_5 e^{-a_6 r}}{r}.
\end{eqnarray}
The $a_i$'s are obtained by fitting  the self-interaction-free DFT potential \cite{Tong97}. 
The other,  which we call here model potential 2, is a two-parameter potential in the following form \cite{Green69,Green73}  
\begin{eqnarray}\label{eq:mp2}
V(r)&=&-\frac{(Z-1)[(\eta/\xi)(e^{\xi r} -1)+1]^{-1}+1}{r}, 
\end{eqnarray} 
where $Z$ is atomic number and the parameters $\eta,\xi$ are obtained by fitting the Hartree-Fock potential \cite{Garvey75}. 
Both model potentials give the correct asymptotic results.

\begin{figure}
\centering{\includegraphics[width=\figwidth]{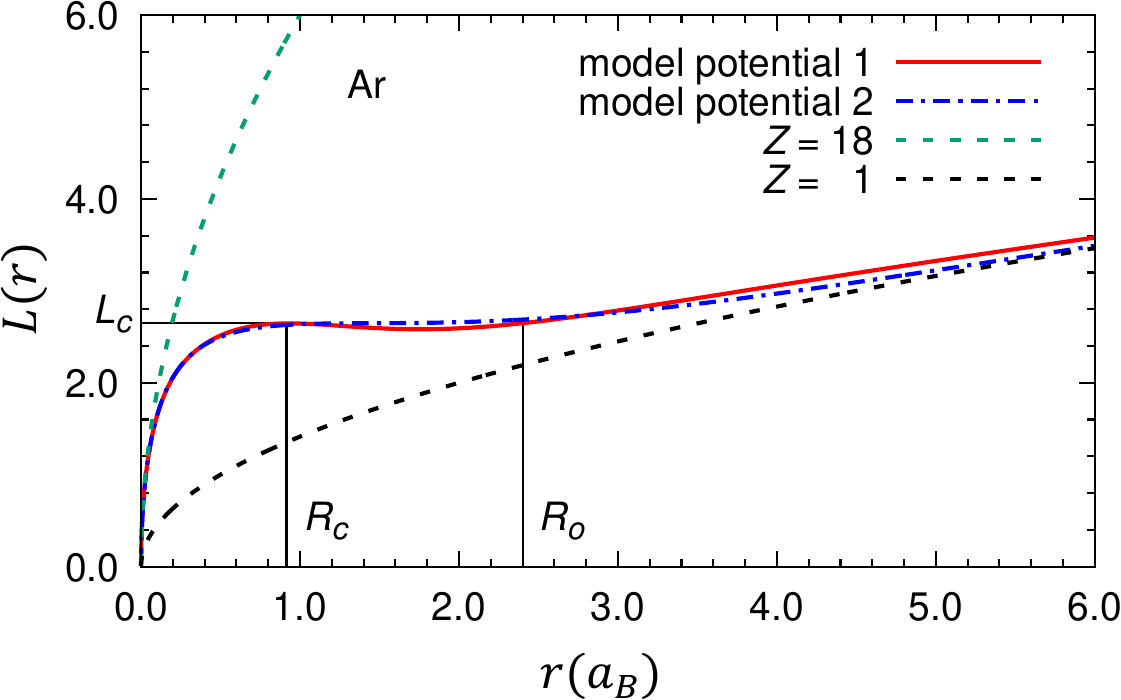}}
\caption{Classical angular momentum $L(r)$ of an electron moving circularly in model potentials of an Ar atom.\label{fig:2}}
\end{figure}

We first analyze the process from a classical viewpoint. If a particle of mass $m$ moves in a radial model potential $V(r)$, a stable circular motion only exists when 
\begin{eqnarray}\label{eq:L}
L^2(r)  = -2mr^2 V(r),
\end{eqnarray} 
where  $L(r)$ is classical angular momentum, a function of the radial position $r$.  We plot  $L(r)$ in Fig.~\ref{fig:2} for the two model potentials, as well as for pure Coulomb potentials  $V(r)=-Z/r$ with $Z=18$ and 1, which represent the two limits when the particle moves in the inner  and outer regions of an Ar atom. 
As $L(r)$ is scaled by $\sqrt{m}$, we set $m = 1$ in Fig.~\ref{fig:2} without loss of generality. 
If  $L(r)$ changes monotonically as a function of $r$ (i.e., for a pure Coulomb potential), there is no orbital collapse. The particle moves in the inner region with  low $L(r)$ and experiences a Coulomb potential of $-18/r$. 
At high $L(r)$, the particle is pushed to the outer region and feels a Coulomb potential of $-1/r$. Figure~\ref{fig:2} shows that as $L(r)$ increases from 0 to 2.6, the stable position moves from 0 to $R_c$ monotonically, and $L(r)$ reaches  the local maximum $L_c$. As $L(r)$ exceeds $L_c$ slightly, the stable position jumps to $R_0$. 
There is no stable circular motion between $R_c$ and $R_0$. As $L(r)$ increases further, the stable position increases monotonically again. 
The circular motion collapses between $R_c$ and $R_0$. Therefore, even for low-$Z$ atoms such as Ar $(Z=18)$  the classical circular orbit may collapse.  

\begin{figure}
\centering{\includegraphics[width=\figwidth]{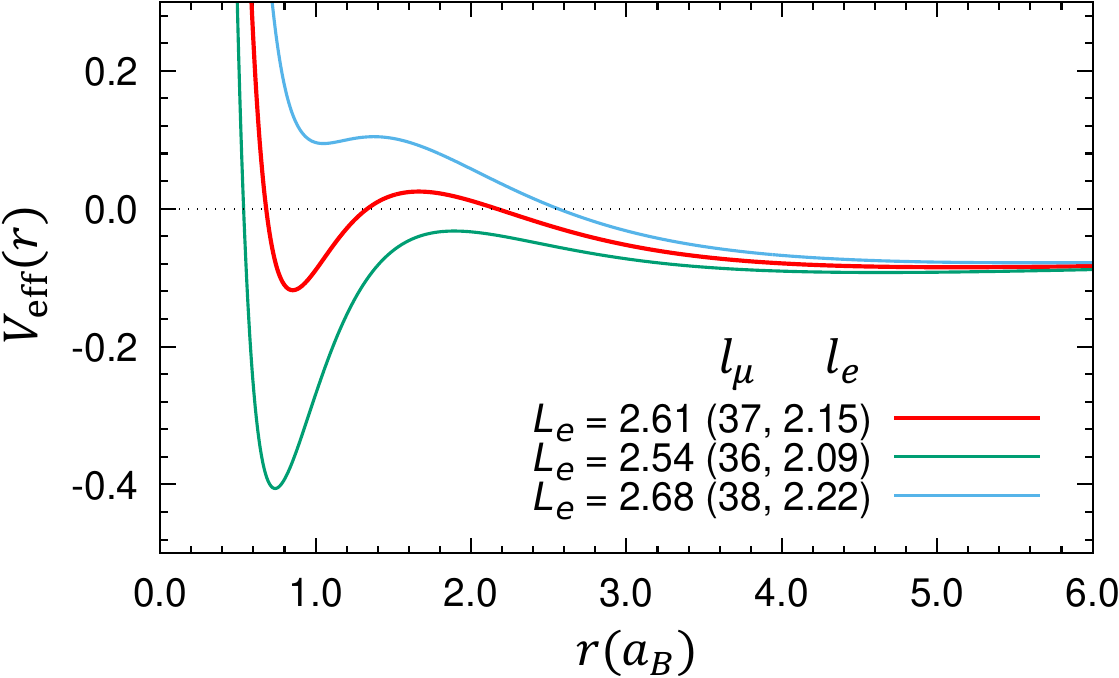} }
\caption{Effective potentials for three classical angular momenta of  $L_e =2.61,2.54,$ and $ 2.68$ using model potential 1. The equivalent  quantum angular momenta of a muon ($l_\mu$) and electron ($l_e$)  are also shown. \label{fig:3}}
\end{figure}

This classical analysis contradicts the lack of orbital collapse for Ar atoms.  Orbital collapse also requires a centrifugal barrier in the effective potential, which is defined as 
\begin{eqnarray}
V_{\mbox{eff}}(r)&  =& \frac{L_e^2}{ 2  r^2} + V(r)\label{eq:ce}\\
&=&\frac{l_e (l_e+1)}{ 2 r^2} + V(r) 
=\frac{l_\mu (l_\mu+1)}{ 2m_\mu r^2} + V(r), \label{eq:ve}
\end{eqnarray}
where $L_e$ is the classical angular momentum of the electron and $l_\mu$ and  $l_e$  are the  quantum angular momenta of a muon and electron, respectively. 
In Eqs.~(\ref{eq:ce}) and (\ref{eq:ve}), $L_e$ and $l_e$ or $l_\mu$ have mathematical one-to-one correspondence,  although non-integer $l_e$ or $l_\mu$ is not allowed in quantum mechanics. We plotted the effective potentials with $l_\mu=36,37,$  and $38$ in Fig.~\ref{fig:3}, and convert $l_\mu$ to $L_e, l_e$.
A significant potential barrier is formed at $L_e=2.61$, which corresponds to $l_e=2.15$ or $l_\mu=37$. The $l_e$ value of $2.15$ is in between the $d$- and $f$-partial waves for an electron, and the value is forbidden  for Ar. This explains the absence of atomic orbital collapse in Ar atoms. However, $l_\mu=l_c=37$ is allowed  for muonic Ar atoms, and this causes the muon orbital collapse at $l_c$.  Note that for $l_\mu=36$, the inner potential well is deep enough to hold a bound state while for $l_\mu=37$ there is no bound state in the inner potential well. If  $L_e$ exceeds  2.61,  the barrier is quickly suppressed. 

Comparing Eqs.~(\ref{eq:ce}) and (\ref{eq:ve}),  we can estimate $l_c$ at which the collapse happens  using $L_c$, the local maximum in Fig.~\ref{fig:2}, as
 \begin{eqnarray}{\label{eq:sc}}
 l_c (l_c+1)=m_\mu L_c^2
\end{eqnarray}
The calculated $l_c$ should be rounded to the nearest integer. This method gives  $l_c=37$.
This implies an important property of heavy particles: due to their heavy mass, muons can probe even a modest change in the effective potential.

As a result of  orbital collapse, the probability of a muon being captured into a state with $l_\mu > l_c$ should be negligible, because there is almost no overlap between the orbitals of the muon and active electron.  To confirm this, we calculated the muon  capture probabilities by the CTMC method with model potential 2 for $E_c = 10$ eV incident energy.   In the simulations,  the active electron $3p$  and moun move in the same model potential,  and the two particles interact  through Coulomb force. 

\begin{figure}[tb]
\centering{\includegraphics[width=\figwidth]{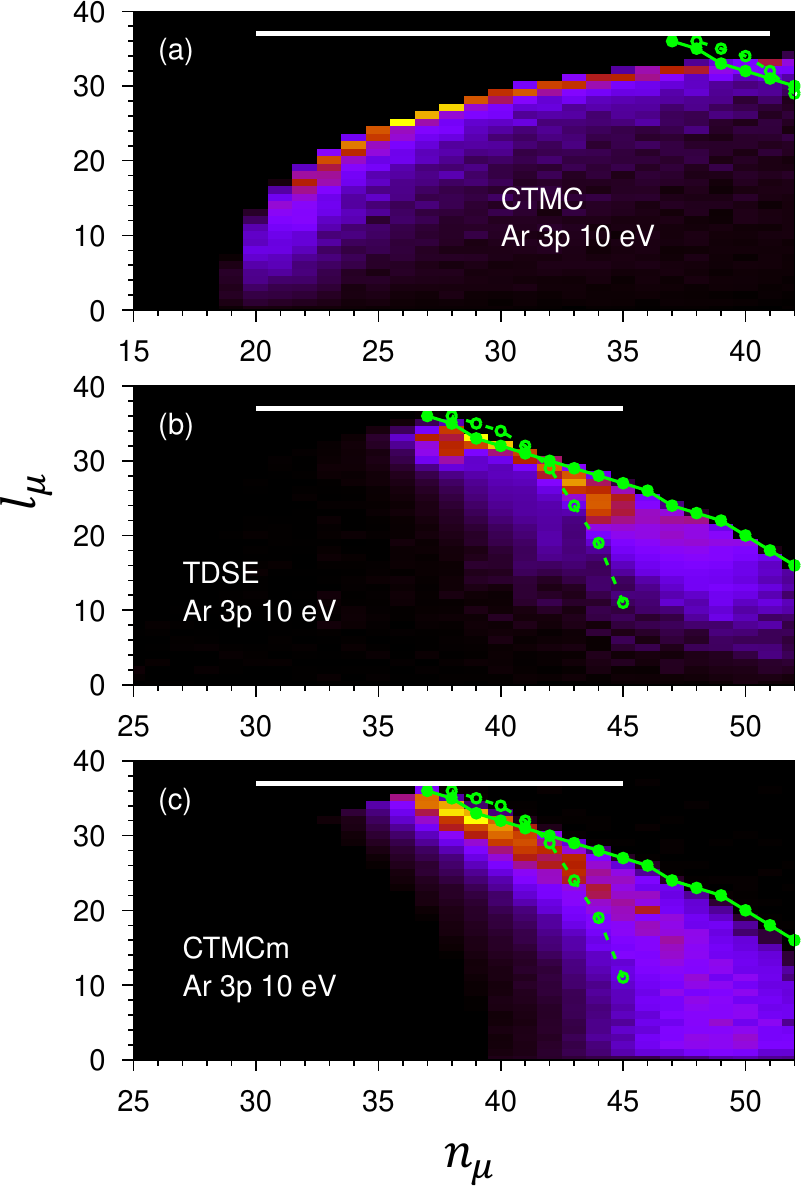} }
\caption{State-specified muon capture probabilities in muon-Ar collisions calculated by (a)  CTMC, (b) TDSE, and (c) the modified CTMC.  The results are normalized to the largest capture probability.  $n_\mu$ in (a) is shifted by 10. Horizontal lines denote $l_c=37$. 
States with a large overlap (open circles) or long interaction time (solid circles)  are presented (see text for details).
 \label{fig:4}}
\end{figure}
The results are plotted in Fig.~\ref{fig:4}(a). The capture probabilities into the states of  $l_\mu \geq l_c$  are negligibly small owing to muon orbital collapse. As $l_\mu$ decreases, the muon is mainly captured into  low-$n_\mu$ states. This tendency agrees with the cases of muon captured by H atoms \cite{Tong07b} and antiproton captured by He atoms \cite{Tokesi05}. The peak position ($n_\mu=25$) is lower than suggested by the recent experiment ($n_\mu$=37) on muonic Ar atoms \cite{Okumura24}. 

To verify the results, we simulated the same process by solving the TDSE \cite{Tong06c,Tong07b} with model potential 1. The results in Fig.~\ref{fig:4}(b) show that the capture probabilities into the states of   $l_\mu \geq l_c$ are negligibly small. As $l_\mu$ decreases, the muon is mostly captured into high-$n_\mu$ states, which contrasts with the results from  CTMC. 
A TDSE simulation with model potential 2 ruled out the discrepancies being associated with the different model potentials.

To understand the discrepancies, we plot the $n_\mu l_\mu$-curve for the highest muon orbital (solid circles) for each $l_\mu$ from energy  conservation in the figure. In such a situation, both the ejected electron and muon move slowly, which increases the interaction time between them and thereby may increases  the capture probability. On the other hand, for a high-$n_\mu$ state, the overlap with the $3p$ electron orbital is reduced.  We  plot the $n_\mu l_\mu$-curve for the state  (open circles), which  mean radius  is most close to the mean radius of the $3p$ electron orbital for each $l_\mu$. The competition between the two generates the present pattern. Thus, we conclude that the TDSE results are more reasonable based on the discussion.

To explore the origin of the discrepancies between Fig.~\ref{fig:4}(a) and Fig.~\ref{fig:4}(b), we examined the details of the CTMC procedures and results. In the CTMC simulation, the binding energy and angular momentum are well defined in classical mechanics. Mapping them to quantum $(n_\mu,l_\mu)$ showed that the same $(n_\mu,l_\mu)$ may correspond to different binding energies for different events in the mapping scheme \cite{Rakovic01,Tokesi05}. To remedy this fatal problem, we calculated the muon orbital energy $\epsilon_{n_\mu l_\mu}$ by solving the following  time-independent Schr\"odinger equation
\begin{eqnarray}\label{eq:2}
\left[-{1\over 2 m_\mu}  {d^2\over dr^2}+ {l_\mu(l_\mu+1)\over 2 m_\mu r^2 } +V(r)\right]\psi_{n_\mu l_\mu} =\epsilon_{n_\mu l_\mu} \psi_{n_\mu l_\mu},\nonumber\\
\end{eqnarray}
with model potential 2.  The capture probability of binding energy $\epsilon_i$ and angular momentum $l_i$ in the CTMC simulation is mapped to $n_\mu,l_\mu$ state as
\begin{eqnarray}
P(n_\mu,l_\mu)&=&\sum_i P(\epsilon_i,l_i)  \ \ 
\mbox{if} \hspace{2mm} l_{-}  < l_i < l_{+}; \ \epsilon_{-} <\epsilon_i < \epsilon_{+}, \nonumber \\
\end{eqnarray}
with $l_{\mp}=l_\mu \mp0.5$ and $ \epsilon_{\mp}=(\epsilon_{n_\mu l_\mu}+\epsilon_{n_\mu \mp 1\ l_\mu})/2$.
We refer to this mapping method as CTMCm. As shown in  Fig.~\ref{fig:4}(c), the  CTMCm results now agree reasonably with the TDSE results. However, the agreement  is still not perfect,  as the problem is complicated. 

\begin{figure}[t]
\centering{\includegraphics[width=\figwidth]{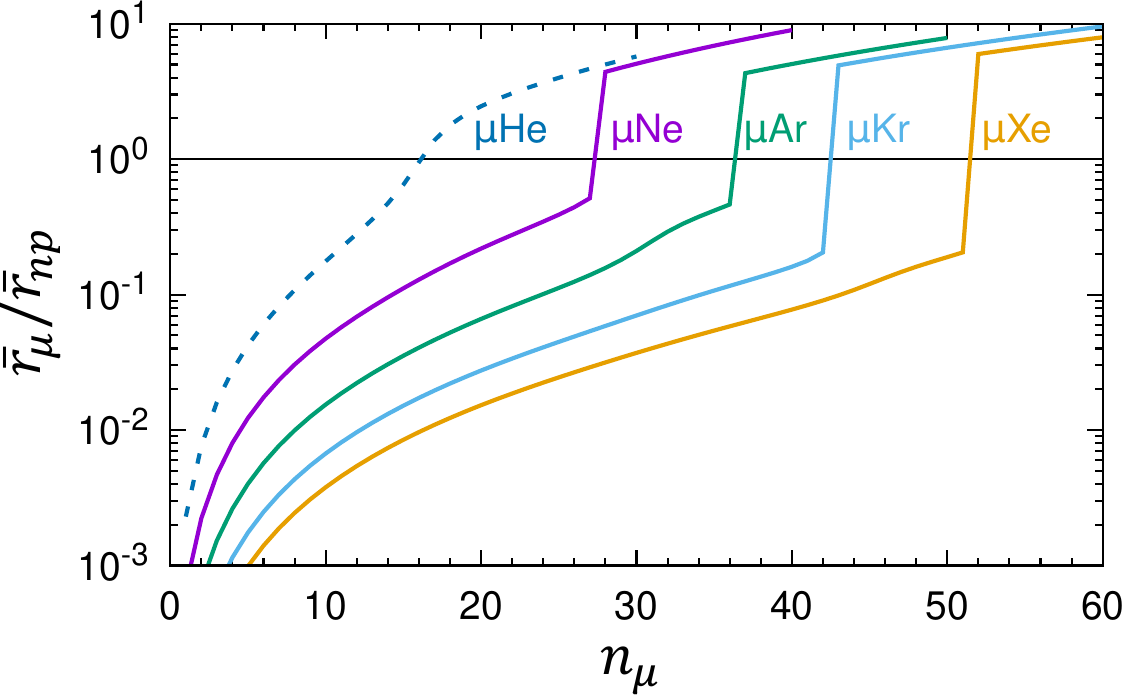} }
\caption{Ratio of the mean radius of muon in a circular state and electron in the highest occupied state for muonic noble atoms.
\label{fig:5}}
\end{figure}
We confirmed muon orbital collapse even for Ar atoms, a low-$Z$  atom at a critical angular momentum $l_c$. The $l_c$ provides the upper limit for the muon capture process. We next considered whether this collapse is a general behavior for other muonic  or other exotic atoms. Fig.~\ref{fig:5}  shows the ratio of the mean radius $\bar{r}_\mu$ of the muon in a circular state and  the mean radius  $\bar{r}_{np}$ of the highest occupied electron $np$ orbital for muonic noble atoms.  Except for muonic He, all the other muonic  noble atoms show a dramatic change in the radial wavefunction. The changes are large for high-$Z$ atoms, and the corresponding $n_\mu$ increases as $Z$ increases.  Indeed, for a muon, orbital collapse occurs  even in muonic Ne atoms. This prediction agrees with the implication of the experiment \cite{Kirch99} that the initial $n_\mu$ exceeds 20 for muonic Ne atoms.

Another consideration is whether  the scaling relation, Eq.~(\ref{eq:sc}), also works for other exotic atom.  Table~\ref{tab:lc} lists the $L_c$ for noble atoms, as obtained using  model potential 1,  $l_c$ from Eq.~(\ref{eq:sc}), and the values from the DFT simulation.
\begin{table}[t]
\caption{$L_c$ of noble atoms from model potential 1 and the critical angular momentum $l_c$ calculated by  DFT or scaled by Eq.~(\ref{eq:sc}) 
for muonic and antiprotonic  atoms. \label{tab:lc}}
\begin{ruledtabular}
\begin{tabular}{lrrrr}
 Atom &   Ne & Ar & Kr &   Xe \\  \hline
$L_c$    				&  1.98   &  2.65   & 3.03  & 3.53  \\
$l_c$ ($\mu$: scaled)   		&   27     	& 37  	&  43       & 50 \\ 
$l_c$ ($\mu$: DFT)    		&   27     	&   36  	&   42     & 51 \\
$l_c$ ($\bar{p}$: scaled) 	&   84    	& 112  	&  129    	&  150  \\  
$l_c$ ($\bar{p}$: DFT)   	&    82	&  109	&  125	&   153 \\  
\end{tabular}
\end{ruledtabular}
\end{table}
 The  $l_c$s for muonic and antiprotonic atoms  from the mass scaling relation and the DFT calculation are similar but not  identical, as the applied model potential does not consider  dynamic changes of the electron wavefunction when the heavy particle locates in different orbitals, as shown in Fig.~\ref{fig:1}. Nevertheless, the analysis based on model potential does predict orbital collapse.

In summary, we studied the energy structures of muonic Ar atoms and found muon orbital collapse at a critical angular momentum $l_c$. This orbital collapse can be explained in terms of the competition among the muon nuclear interaction, the electron screening effect, and the muon centrifugal potential in a circular state. The parameter $l_c$ provides an upper limit for the angular momentum of the muon-captured states. We confirmed the existence of the upper limit using the time-dependent Schr\"odinger equation and classical trajectory Monte Carlo simulations.  Furthermore, by modifying the mapping between the classical energy and the principal quantum number,  we attained a reasonable agreement for the state-specified capture probabilities obtained by the two methods. We also showed that orbital collapse is a general phenomenon for other exotic atoms, even for muonic Ne. The present work may stimulate other studies related to orbital collapse, such as  the giant resonance in the photoionization \cite{Amusia89,Tong00b,Takahashi07}, Auger decay involved in the collapsed orbital \cite{Zhong20}. 
\\

All data are available from the corresponding author (X.M.T.) upon request.

\begin{acknowledgments}
This work was supported by the Multidisciplinary Cooperative Research Program in Center for Computational
Sciences, University of Tsukuba, the NIFS collaboration program (NIFS23KIIF031),  and RIKEN Pioneering Projects. X.M.T. was supported by
JSPS KAKENHI (Grant-in-Aid for Scientific Research
C) 22K03493.

\end{acknowledgments}

\input Orbital-collapse-PRL-v0.bbl

\end{document}

%% file: Orbital-collapse-PRL-v0.bbl
%